\documentstyle[12pt]{article}
\def\be{\begin{equation}}
\def\ee{\end{equation}}
\def\ba{\begin{eqnarray}}
\def\ea{\end{eqnarray}}
\def\lsim{\mathrel{\mathop
  {\hbox{\lower0.5ex\hbox{$\sim$}\kern-0.8em\lower-0.7ex\hbox{$<$}}}}}
\def\gsim{\mathrel{\mathop
  {\hbox{\lower0.5ex\hbox{$\sim$}\kern-0.8em\lower-0.7ex\hbox{$>$}}}}}
\def\bq{\begin{quote}}
\def\eq{\end{quote}}

 at 10truept

\newcommand{\ii}{{\'\i}}

\newcommand{\beq}{\begin{equation}}
\newcommand{\eeq}{\end{equation}}
\newcommand{\beqa}{\begin{eqnarray}}
\newcommand{\eeqa}{\end{eqnarray}}

\def\la{~\mbox{\raisebox{-.6ex}{$\stackrel{<}{\sim}$}}~}

\def\ltap{\ \raise.3ex\hbox{$<$\kern-.75em\lower1ex\hbox{$\sim$}}\ }
\def\gtap{\ \raise.3ex\hbox{$>$\kern-.75em\lower1ex\hbox{$\sim$}}\ }
\def\gl{\ \raise.5ex\hbox{$>$}\kern-.8em\lower.5ex\hbox{$<$}\ }
\def\roughly#1{\raise.3ex\hbox{$#1$\kern-.75em\lower1ex\hbox{$\sim$}}}

\begin{document}
\thispagestyle{empty}
\begin{flushright}
IFT-UAM/CSIC-02-56\\ hep-th/0212132\\ December 2002
\end{flushright}
\vspace*{1cm}
\begin{center}
{\Large \bf Black Hole Astrophysics in AdS Braneworlds }\\
\vspace*{.7cm}
{\large Roberto Emparan$^{a,}$\footnote[1]{{\tt
 roberto.emparan@cern.ch}},
Juan Garc\ii a-Bellido$^{b,}$\footnote[2]{{\tt bellido@mail.cern.ch}}
and Nemanja Kaloper$^{c,}$\footnote[3]{\tt kaloper@physics.ucdavis.edu}}\\[6mm]
{\em $^a$ Theory Division, CERN, CH-1211 Geneva 23, Switzerland}\\[3mm]
{\em $^b$ Departamento de F\'\i sica Te\'orica \ C-XI, Universidad
Aut\'onoma de Madrid, Cantoblanco, 28049 Madrid, Spain}\\[3mm]
{\em $^c$ Department of Physics, University of California, Davis,
CA 95616, USA}\\

\vspace{1.5cm} ABSTRACT
\end{center}
We consider astrophysics of large black holes localized on the brane in
the infinite Randall-Sundrum model. Using their description in terms of
a conformal field theory (CFT) coupled to gravity, deduced in Ref. [1],
we show that they undergo a period of rapid decay via Hawking radiation
of CFT modes. For example, a black hole of mass ${\rm few} \times
M_\odot$ would shed most of its mass in $\sim 10^4 - 10^5$ years if the
AdS radius is $L \sim 10^{-1}$ mm, currently the upper bound from
table-top experiments. Since this is within the mass range of X-ray binary
systems containing a black hole, the evaporation enhanced by the hidden
sector CFT modes could cause the disappearance of X-ray sources on the
sky. This would be a striking signature of RS2 with a large AdS
radius. Alternatively, for shorter AdS radii, the evaporation would be
slower. In such cases, the persistence of X-ray binaries with black
holes already implies an upper bound on the AdS radius of $L \ltap
10^{-2}$ mm, an order of magnitude better than the bounds from
table-top experiments. The observation of primordial black holes with a
mass in the MACHO range $M \sim 0.1 - 0.5 ~ M_\odot$ and an age
comparable to the age of the universe would further strengthen the bound
on the AdS radius to $L \ltap {\rm few} \times 10^{-6} $ mm.

\vfill
\setcounter{page}{0}
\setcounter{footnote}{0}
\newpage

\section{ Introduction}

In recent work \cite{efk} two of us (together with A.~Fabbri) have
considered black holes localized on branes in Anti-deSitter (AdS)
braneworlds, as proposed by Randall and Sundrum (RS2) \cite{rs2}.  Our
main result was deducing a connection between the classical bulk
dynamics of black holes localized on branes in AdS$_{D+1}$ and the
semiclassical description of black holes in the dual CFT+gravity theory
in one dimension fewer\footnote{The basic rules of this duality have
been considered in, for example, \cite{apr1,apr2,apr7}.}.  This
connection is summarized in our conjecture \cite{efk} that {\it the
black hole solutions localized on the brane, found by solving the
classical bulk equations in} AdS$_{D+1}$ {\it with brane boundary
conditions, correspond to quantum-corrected black holes in $D$
dimensions, rather than classical ones.} This follows naturally from the
AdS/CFT correspondence adapted to AdS braneworlds.  Ref.~\cite{efk}
provides a substantial amount of evidence supporting this result.

An immediate application of this conjecture determines the
evolution of black holes in the RS2 model. The evolution of black
holes larger than the AdS radius $L$ can be fully described either
by the classical bulk dynamics or by the semiclassical large $N$
dynamics of the dual CFT+gravity. Resorting to the latter, we can
see that the presence of a large number of CFT degrees of freedom
accelerates the decay of a black hole via Hawking radiation.
This has also been noticed in Ref.~\cite{tanaka}.

This is qualitatively different from black hole evaporation in
braneworld models where the curvature of extra dimensions can be
ignored, as studied in \cite{ehm3}. Our picture applies only to
models which admit a dual CFT+gravity description, which exists
when the bulk is asymptotically AdS as in RS2 braneworlds (unlike,
e.g. ADD scenarios \cite{add}). Further, black holes must be {\it
larger} than the asymptotic AdS radius $L$. Then, as long as the
CFT+gravity dual is valid, regardless of how cold a black hole is,
it can always access a large number of CFT modes, because the
asymptotic AdS geometry in the bulk means that there is no gap to
suppress the CFT emission. From the bulk point of view, this
effect is a {\it classical} one, which arises because a black hole
stuck on a brane is accelerating through the AdS bulk. The
classical emission of bulk radiation by these large black holes
will overshadow the quantum radiation considered in \cite{ehm3}.
The radiation we
are considering only looks quantum once one turns to the dual CFT+gravity,
where the largesse of this effect compared to the emission of brane modes
is due to the huge number of CFT degrees of freedom -- which is in turn
the dual of the huge ratio $(L/L_{P4})^2$.

Black holes which are smaller than the AdS radius should radiate
more slowly. Indeed, the dual CFT+gravity description breaks down,
because the bulk graviton modes are needed to describe the black
hole geometry itself and cannot be interpreted as dual CFT degrees
of freedom \cite{efk}. The evolution of small black holes can
still be described by a semiclassical bulk theory without evoking
the dual CFT+gravity. This description is very sensitive to the
details of the UV completion of the bulk theory. Still, one sees
from it that the rate of Hawking evaporation diminishes
significantly, as suggested by the classical bulk picture
argument. A small black hole Hawking-radiates mostly along the
brane, since the dominant {\it s}-wave channel cannot discriminate
between the bulk and the localized gravity \cite{ehm3}. This evaporation
is much slower than the emission of a large number of CFT modes.

A complete picture of the evolution of black holes in RS2
braneworlds which start out larger than the AdS radius  therefore
comprises of two stages: a period of rapid mass loss, during which
the CFT+gravity description remains valid, and a period of slow
evaporation, during which the black hole continues to lose mass at
a much lower rate. The latter stage can last a very long time if
the AdS radius is close to the observational bounds. However,
these black holes are so small that their observational signatures
are weak and could only become important due to the cumulative
effect of exciting too much relativistic CFT after inflation. We
will therefore focus on the effects of large black holes in what
follows.

The most dramatic observable consequence of the classical picture
of black hole evaporation is that the black hole lifetime can be
tremendously shortened. In this article we will exploit the
consequences of this observation for two complementary purposes:
(1) deriving bounds on the AdS radius $L$ in RS2 braneworlds from
the observation of astrophysical black holes, and, somewhat more
speculatively, (2) proposing new ways for observing large, or
infinite warped extra dimensions. We find that the black hole
lifetime provides the most sensitive probe of the number of CFT
degrees of freedom. In the former case, they lead to bounds on $L$
which are 3 to 5 orders of magnitude stronger than the table-top
and cosmological limits, which at present constrain $L$ to less
than about $10^{-1}$ mm \cite{eotwash,apr3,hmr}. In the latter
case, they lead to a rapid mass loss of large black holes, which
could deplete the population of super-solar mass black holes in
the universe.

\section{A Qualitative Picture of the Black Hole Evolution}

The evaporation
time of a black hole larger than the AdS$_5$ radius $L$ in the RS2
setup is easily derived using the dual 4D CFT+gravity description.
As a first estimate of their lifetime, to be refined below, recall
that in the presence of quantum fields with a number $g_*$ of
light degrees of freedom, a black hole will evaporate in a time
roughly given by
\be
\tau \sim \frac{10^{64}}{g_*} \left(\frac{M}{M_\odot}\right)^3 \rm{yr}\,.
\ee
The number of CFT degrees of freedom is related to the bulk
AdS radius $L$ by $g_* \sim (L/L_{P4})^2$ where $L_{P4}$ is the
$4D$ Planck length, so $g_*\sim 10^{64} (L/{\rm 1mm})^2$ and
\be
\label{lifet}
\tau \sim \left(\frac{M}{M_\odot}\right)^3
\left(\frac{\rm 1mm}{L}\right)^2 ~\rm{yr}\,.
\ee
Hence, as described above, the decay is extremely rapid when compared
to more conventional cases, such as the Standard Model (SM) coupled
to $4D$ gravity, with a number of matter degrees of freedom $\sim
{\cal O}(100)$ \cite{swh,page:76}. For example, in the case of
SM+gravity it would take an age of the universe for a black hole
the size of Mount Everest, of mass $M \sim 10^{15}$ g, and size
$R\sim 10^{-15}$ m, to completely disappear \cite{swh}, whereas in
RS2 with $L \ltap 0.1$ mm, a black hole of a few solar masses
might evaporate most of its mass in a time of the order of
millenia. Note that our approach is valid for $G_4 M \gtap L$, which
is satisfied by most astrophysically relevant black holes. Indeed,
\be
\label{masslim}
M \gtap 10^{-6} \left(\frac{L}{\rm 1mm}\right) M_\odot
\simeq 10^{51} \left(\frac{L}{\rm 1mm}\right) {\rm GeV}\,,
\ee
so we see that, even with the current upper bound from table-top
experiments limit \cite{eotwash}, $L \ltap 0.1$ mm, the dual
CFT+gravity description of black holes applies to most of the
interesting cases for astrophysical considerations, the more so
the smaller $L$ is. Thus, black holes in the range
(\ref{masslim}), including quite heavy ones,
would evaporate away most of their mass very
quickly. If this process actually takes place, it can
significantly deplete the population of large, solar mass black
holes in the universe. Hence, observations of such black holes
would imply the absence of this effect
and enable us to derive an upper
bound on the number of CFT degrees of freedom $g_*$, which
translates into a bound on the bulk AdS radius $L$. Alternatively,
one might under special circumstances end up with a situation
in which all the observed black holes come from black hole
progenitors that have managed to accrete enough mass from the
environment to compensate for the evaporation. In this case, one could
search for the disappearance of X-ray sources on the sky,
corresponding to the transition of a black hole from being
larger to being smaller than the AdS radius, i.e. smaller
than the bound (\ref{masslim}).

We note that recently an attempt was made to study the evolution
of primordial black holes (PBHs) in RS2 \cite{cgl}. However
\cite{cgl} does not take into account the effects we study here,
and therefore misses the most relevant physics for black holes
larger than the AdS radius. We now turn to the more precise
determination of the evaporation time.

\section{The Evaporation Time of Large Black Holes}

We now make formula (\ref{lifet})
more precise. The rate of mass loss for a neutral, non-rotating
black hole in vacuum due to Hawking evaporation is
\be
\frac{dM}{dt}=-\frac{\alpha}{(G_4M)^2}
\ee
where $\alpha$ is a factor that accounts for the Stefan-Boltzmann
factors and greybody corrections for the particles of each spin that are
emitted,
\be
\alpha=n_0\;\alpha_0+n_{1/2}\;\alpha_{1/2}+n_1\;\alpha_1 \, .
\ee
Here $n_0$, $n_{1/2}$ and $n_1$ are the numbers of real scalars, Weyl
(or Majorana) fermions, and gauge vectors, respectively.  The
coefficients $\alpha_s$ have been computed as\footnote{$\alpha_{1/2}$
and $\alpha_{1}$ were computed numerically in \cite{page:76}, while
$\alpha_0$ was obtained in the optical approximation in
\cite{dewitt}. See also \cite{vaz}.}
\be
\alpha_0=7.0\times 10^{-5},\quad \alpha_{1/2}=8.0\times 10^{-5},\quad
\alpha_1=3.4\times 10^{-5}\,.
\ee

The specific matter content, $n_s$, of the dual CFT depends on the
details of the UV completion of the theory, e.g. on how the model is
embedded into string theory (or any consistent quantum gravity theory).
In the absence of a concrete embedding, we will use the simplest case
where the full space is the AdS$_5\times S^5$ background of IIB string
theory. This is dual to ${\cal N}=4$ supersymmetric $SU(N)$ Yang-Mills
theory, which for a large $N$ yields $n_0=6N^2$, $n_{1/2}=4N^2$ and
$n_1=N^2$. On the other hand, AdS/CFT gives $N^2=\pi L^3/2G_5$
\cite{adscft}, while RS2 implies $G_5= L G_4$, so $N^2=\pi L^2/2G_4$.
Thus one obtains
\be\label{lossrate}
\frac{dM}{dt}=-2.8\times
10^{-3}\left(\frac{M_\odot}{M}\right)^2\left(\frac{L}{\rm 1mm}\right)^2
M_\odot ~\rm{yr}^{-1} \, .
\ee
Integrating this equation and dropping the corrections of order of the critical
mass of Eq. (\ref{masslim}), gives the time within which a large black
hole initially of a mass $M$ will shrink to below the size of $L$:
\be\label{lifet2}
\tau \simeq 1.2\times 10^{2}\left(\frac{M}{M_\odot}\right)^3
 \left(\frac{\rm 1mm}{L}\right)^2 ~\rm{yr}\,,
\ee
which is larger than the estimate (\ref{lifet}), that neglected
greybody corrections and other factors. From Eq.~(\ref{lifet2}), we note
that in the extreme case where $L\simeq 0.1$ mm, a black hole of a mass $ M
\ltap 50$ $M_\odot$ which could have formed by the collapse of early
stars would have evaporated by today down to a small mass below the bound of
Eq. (\ref{masslim}). Thus the strongest bounds and signatures will come from either
primordial or stellar-mass black holes.

The assumption that the matter content of the dual CFT is that of ${\cal
N}=4$ $SU(N)$ SYM is not essential. For other embeddings of AdS$_5$ the
final result Eqs.~(\ref{lossrate}) and (\ref{lifet2}) may be modified
only by factors of order unity: whenever the bulk geometry is asymptotically
AdS$_5$, the dual theory will contain a large number $\sim
(L/L_{P4})^2$ of light fields. {\it Any} such theory must
reproduce {\it exactly} the same Weyl anomaly, regardless of its matter
content, so the final result must always be quite close to the above.
This picture would not apply if the AdS$_5$ geometry were modified
at some distance away from the brane
such that the dual theory develops a mass gap greater than the black
hole temperature\footnote{This is one of the reasons why our arguments
do not apply to ADD.}. Observe that, from the CFT+gravity viewpoint,
the quickened evaporation rate is entirely due to the vast increase in
the phase space available to black hole emission. The temperature of black
holes that are larger than $L$ is essentially unaffected by the
presence of the extra dimensions, taking
the usual form
\be
T\simeq 6\times 10^{-8}\frac{M_\odot}{M}\; {\rm K}\,,
\ee
which is extremely small for astrophysical black holes. However, if
the bulk is modified far from the brane, the dual theory has a gap
and ceases to be a CFT in the IR. The gap would suppress the emission
of Hawking radiation from black holes
whose temperature is below it. In such instances our
constraints could be significantly weakened, depending on
the scale of the gap.

The main practical obstacle for deriving the bounds comes from the
fact that realistic black holes do not live in a vacuum. They are
surrounded by a distribution of matter, such as the thermal
background photons and the interstellar medium for isolated black
holes, and even more importantly, the dense environment for black
holes inside galaxies. Examples are the gas of a companion star in
the case of X-ray binaries, which typically form accretion disks,
like in microquasars \cite{microquasars}, and the halo of ionized
infalling matter stripped from nearby stars, in the case of large
quiescent black holes in the centers of galaxies. Any black hole
colder than the environment will absorb energy from its
surroundings and accrete mass onto itself \cite{hawkcar}. The rate
at which a black hole accretes mass depends very much on its
environment and ranges from $10^{-9}\ M_\odot$ yr$^{-1}$, for
isolated black holes and low mass X-ray binaries, to $ M_\odot$
yr$^{-1}$ for high redshift quasars.

The processes of evaporation and accretion compete with each
other. However they work quite differently: while the evaporation
rate decreases with the black hole mass as $M^{-2}$, the accretion
grows proportionally to the horizon area and hence to $M^2$. The
dynamical law describing the rate of mass change of a
black hole with accretion included is \cite{hawkcar}
\be
\dot M = \beta (G_4 M)^2\,\rho - \frac{\alpha}{(G_4 M)^2} \, ,
\label{masschange}
\ee
where the first term encodes the accretion
of matter with energy density $\rho$ surrounding the black hole
and the second term encodes Hawking evaporation (\ref{lossrate}).
Here $\beta$ is a numerical coefficient measuring the efficiency
of accretion, which may depend on the velocity of the black hole
moving through the intergalactic medium.
Ignoring details we will assume that it is of order
unity. Now, the accretion will be smaller when the evaporation
proceeds faster, i.e. when the black hole is smaller.  From
(\ref{masschange}) we see that this will happen when $\rho (G_4
M)^4 < 10^{64} (L/1{\rm mm})^2$.  For example, massive black
holes in the core of AGNs have accretion rates\footnote{Estimated
from their observed luminosity, see e.g. Ref.~\cite{accretion}.} of
order $1- 10^{-4} M_\odot {\rm yr^{-1}}$, which completely
overwhelm the evaporation rate (\ref{lossrate}).

For stellar mass black holes the situation is more complicated,
but it is expected that the typical rate at which a normal star is
dumping gas onto a companion black hole is around $10^{-9} M_\odot
{\rm yr^{-1}}$. This gives a rather wide margin for the evaporation
to dominate. Thus, isolated or smaller black holes, while harder
to detect, will end up being dominated by the evaporation into CFT
modes. In this instance, it is straightforward to check that the
accretion of CMB photons is negligible, even if the CMB is much
hotter than the black hole. For example, for a solar mass black
hole in theories where $L\sim 0.1$ mm, the mass increase due to
the accretion of CMB photons at temperatures below the MeV is
negligible compared to the mass loss coming from CFT Hawking evaporation.
Thus even primordial black holes will end up mostly evaporating
CFT throughout their lifetime. This allows us to place bounds on
$L$, and also argue for new ways of looking for the signatures of RS2.
We now turn to the best sources of the bounds.

\section{The Bounds on {\it L}}

~

\indent{\bf Sub-solar-mass black holes and MACHOs.}
\medskip

It is clear from the arguments above that CFT evaporation
affects most dramatically smaller black holes, whose horizon exceeds
the AdS radius $L$ just barely, such that they are the hottest black holes
allowing dual CFT+gravity description. Specifically, the
sub-solar-mass black holes will yield the strongest bounds. They could
not have formed by collapse of matter because of the Oppenheimer-Volkoff
bounds, and hence must be primordial, forming during some violent process
in the very early universe~\cite{gbwl,yoko,jedam}. Such black holes must have
lived at least as long as the age of the Universe, $\tau_U\sim 14$ Gyr.
An observation of a single such black hole would thus strongly
constrain the parameter $L$. From this $\tau_U$ and Eq. (\ref{lifet2})
we obtain
\be
L \la 10^{-4}\left(\frac{M}{M_\odot}\right)^{3/2} {\rm mm} \,.
\ee
Hence any evidence for a small black hole with mass
in the range $M \sim 0.1 -1 M_\odot$ would impose a bound on $L$ that
may be as tight as
\be
L \la {\rm few} \times 10^{-6} {\rm mm} \, ,
\ee
which is some five orders of magnitude stronger than the current limits
from table-top experiments and cosmology~\cite{eotwash,apr3,hmr}.

What are the chances of observing such black holes? It has been
argued recently that they might actually comprise a significant
portion of the dark matter in the galaxy halo, to avoid problems
with having too many baryons if the MACHOs were all brown dwarfs
\cite{ffg}. If this is true, the most obvious difficulty for their
detection is to distinguish the PBH MACHOs from other compact,
nonluminous objects in the {\it same} mass range. At present, the
best strategy for identifying such light objects as black holes is
to search for the gravitational waves emitted during the merger
phase of a coalescing binary containing a black hole \cite{nstt}.
This challenge has apparently been taken by the GEO-600
collaboration, which has plans to search for binaries containing
black holes in the MACHO range \cite{schutz}.

\medskip
{\bf Isolated few-solar-mass black holes.}
\medskip

CFT emission could have played a
significant role for black holes with a mass in the range $1-100
M_\odot$, formed by stellar collapse, if they were old enough.
If the AdS radius $L$ is sufficiently
large, say, more than a fraction of a micron, this will be the
case for black holes formed in the supernova explosion of
first-generation stars, $t \sim 5\times 10^9$ yr ago. An
additional requirement is that the black hole be isolated, so it
does not accrete from a companion star.  While their
detection is difficult, it is possible in principle, through the
emission of X-rays via accretion of gas from the interstellar
medium (ISM), see Ref. \cite{Agol:2001hb}. It has been argued by
Brown and Bethe \cite{BB} that there must be a
large population ($\sim 10^9$) of black holes in our galaxy with a mass around
$1.5 M_\odot$. If so, there should be many microlensing
events with such a mass towards the galactic bulge.
In the future such black hole candidates spotted with microlensing can
be checked with X-ray measurements by XMM and Chandra, which could
give crucial information about the low-mass black holes in the
galaxy. As a first step, recently both MACHO and OGLE
collaborations have claimed discovery of
nearby black holes through long-lasting microlensing events
\cite{Bennett:2001vh,Mao:2001xs}.

Isolated black holes are expected to accrete ISM matter at a
very low rate \cite{Agol:2001hb}, $\dot M_{\rm accr} \sim
1.5\times10^{-11} \,M_\odot {\rm yr^{-1}}$, and therefore their rate of
evaporation into CFT modes (\ref{lossrate}) would be significantly
larger $|\dot M_{\rm evap}| \sim 1.2\times10^{-3}\,(L/1{\rm
mm})^2\,M_\odot {\rm yr^{-1}}$, unless the AdS radius
is $L < 10^{-4}$ mm. In this case a few-solar-mass black hole could retain
most of its mass for few billion years. Verifying
the existence of a population of black holes with masses of few $M_\odot$
would therefore impose a stringent bound on $L$,
\be
L \la {\rm few} \times 10^{-4} {\rm mm} \, ,
\ee
about three orders of magnitude stronger than the limits from
table-top experiments \cite{eotwash}.

\medskip
{\bf Black holes in X-ray binaries.}
\medskip

The best observational candidates for black holes are X-ray
binaries. They come in two classes, low mass (LMXB) and high mass
(HMXB) X-ray binaries, depending on the mass of the companion,
which can be determined reliably from the orbital properties of the
companion in a binary system. Most of the best known black holes,
with masses measured to better than 10\% error, correspond to LMXB
and, more precisely, to the so called Soft X-ray Transients, also
known as {\em X-ray Novae}~\cite{Casares}. A dozen of them are
known to date, with their orbital properties and masses detailed
in Table 1. of Ref. \cite{BHbinaries}. Their average mass is
$\langle M\rangle \sim 7\,M_\odot$, but the dispersion is large,
since the masses range
from $3\,M_\odot$ to $14\,M_\odot$. Most of these LMXB
were formed when the heavy star of the binary exploded in a
supernova, without completely blowing the system apart, and
leaving behind a black hole. Their typical accretion rates from their
companion stars are of order of $\dot M_{\rm accr} =
 10^{-10}\,M_\odot {\rm yr^{-1}}$~\cite{Casares}. This
accretion rate sets a benchmark with which to compare the emission
of Hawking radiation in the CFT sector.

Unfortunately, less is known about the lifetimes of X-ray
binaries. Most black holes in binaries were formed from heavy
stars in the disk of the galaxy, so their lifetime is estimated to
be $t < t_{\rm disk} \sim 10$ Gyr. However, reliable ages are
difficult to determine, and there are very few examples of
reliable estimates. These involve young binary systems, such as
the X-ray transient XTE J1118+480 and the microquasar GRO J1655-40
(also known as Nova Scorpii 1994). The mass of the black hole in XTE J1118+480
is $M = 6.8\pm0.4\, M_\odot$, and its age has been estimated
by using the properties of its eccentric halo orbit which happened
to pass close to Earth. This led to the suggestion that this black hole
may have originated in the galaxy disk 240 Myr ago and went out of
the plane due to a supernova explosion \cite{Mirabel}.
Substituting the mass and the age in Eq. (\ref{lifet2}),  we find
the bound on $L$ to be
\be
\label{imprbd}
L \ltap 1.3\times10^{-2} {\rm mm} \, ,
\ee
already an order of magnitude better than table-top
experiments. A systematic uncertainty in the age of the black hole
towards larger values \cite{Mirabel} could give a somewhat
stronger constraint on $L$. This alone would make it hard for
table-top experiments to detect the subleading corrections to $4D$
gravity in RS2. We note that the age estimate of XTE J1118+480 has
been questioned in \cite{Mirabel,private}, where the possibility
was raised that the black hole may have originated in a globular
cluster 7 Gyr ago and followed a path in the halo characteristic
of globular clusters. This would significantly improve the bound
on the AdS radius, leading to
\be
L \ltap 2.4\times10^{-4} {\rm mm} \, ,
\ee
which is three orders of magnitude better than the
bounds from table-top experiments.

The GRO J1655-40 microquasar has a black hole whose mass is $M =
5.4\pm0.3\,M_\odot$ \cite{Beer}. Its age can be estimated
from a peculiar chemical abundance of the surface of the
companion, with elements that can be traced back to a supernova
explosion of type II. Since the rate at which the companion loses
matter towards the accretion disk is known, one can place the limit
of $\la 10^{6}$ yr on the age of the black hole \cite{Rebolo,Combi}.
Very recently, multiple wavelength observations of GRO J1655-40
were used to determine its velocity and possible origin in the sky
\cite{Mirabel2002}. From these observations, providing
the first direct evidence for stellar black hole formation in
supernovae explosions, one can estimate the age
of the black hole to be 0.7 Myr~\cite{Mirabel2002}.
Altogether this gives only a mild bound on $L \ltap 0.2$ mm, comparable to
table-top experiments.

Future measurements of the dozen or so X-ray novae with XMM and
Chandra satellites may reveal their origin and their age. Any
black hole as old as the galaxy would then impose a strong bound
on the AdS radius, of order
\be
L \ltap {\rm few} \times10^{-3} {\rm mm} \, .
\ee

\medskip
{\bf Improved bounds from rotating black holes.}
\medskip

So far we have been considering evaporating black holes which are
static, without any angular momentum. This is probably
unrealistic: most black holes are expected to be spinning, some
with angular momenta close to the unattainable extremal limit
$a\equiv J^2/M^2=a_{max}=1$ in geometric units. There is some
observational
evidence for black holes in the stellar mass range with spins up
to $a=0.94$, and also of galactic black holes with similarly large
spins \cite{spinbh,otherspin}.

A black hole loses its angular momentum by Hawking emission of
particles with both orbital and intrinsic angular momentum
\cite{page:76b}. Indeed, a quickly rotating black hole is more likely
to emit particles of higher spin, an amplification
closely related to superradiance \cite{superrad,aas,unruh}. Explicit calculations
show that the addition of some spin to the black hole does not, in fact,
change its lifetime by more than a factor of about 2 \cite{page:76b}.
Hence our estimates above which ignored the effects of spin are
still reliable.

However, there remains the possibility that measurements of the spin of
a black hole, rather than of the mass, may yield improved bounds on the
number of CFT modes, and thus on the curvature scale $L$ of the
extra dimension. A quickly spinning black hole, $a\ltap 1$, will
shed off a large fraction of its spin quite fast until it is
reduced to more modest values of $a$. In our context, the enhanced
evaporation rate into CFT modes will amplify this effect and
prevent a black hole from retaining a large angular momentum for
a long time. From the analysis in \cite{page:76b} we infer that a
black hole with an angular momentum above $a=0.9$ will spin-down
below this value in a time equal to around 1/50 of its total
lifetime. Thus, the existence of angular momenta in the range
$0.9\leq a<1$ that have been held for a
sufficiently long time will further improve the
bounds on $L$ by as much as an order of magnitude beyond those obtained from
the total lifetime of the black hole. As an example, if the
candidate spinning black hole of Ref.~\cite{spinbh,otherspin} is
as old as $10^9$ yr, then
\be L<10^{-4} {\rm mm} \, . \ee

\section{Black Hole Detonations as a Signature of RS2}

So far we have been focusing on the bounds on the AdS radius which come
from requiring that black holes with masses $\la {\rm few} \times M_\odot$
evaporate sufficiently slowly to be around today, even if they were formed
billions of years ago. However, the numbers of the observational black
hole candidates in this mass range are in the teens. Thus while it may not
be very likely, it is possible that those which really are black holes
ended up surviving for so long because of accretion, that might have won
over fast CFT emission. This would give rise to an exciting possibility
for very significant signatures of RS2 with as large an AdS radius as
currently allowed by observations, given by our Eq. (\ref{imprbd}), $L
\la$ 0.01 mm.

In this instance, Hawking evaporation of CFT modes from a black hole with
a mass close to a solar mass in a X-ray binary in the last stages of
evaporation would lead to a decay time $\tau$ of order of few million
years, rendering the mass time-dependent, $M(t) = M_{\rm BH}
(1-t/\tau)^{1/3}$. In the final stages of the evaporation into the CFT
sector, the mass of the black hole would change rapidly: it would reduce
to one-tenth of its initial value in the final millenium, and to
one-hundredth in the last year, and so forth. Since these time scales are
much shorter than any dynamical (gravitational collapse) time scales, the
potential well holding the rotating accretion disk together (as well as
the companion star in its orbit) would disappear suddenly, and the disk
would fly apart by the action of centrifugal forces, sending the companion
outward like a slingshot. No events of this kind have been observed yet.
However, if the initial mass function for stellar formation favors the
formation of lower-mass objects, there could be a large
population~\cite{BB} of solar mass black holes ready to go away. An
observation of such a dramatic event would immediately suggest sudden
black hole evaporation. One could also imagine searching for stars that
have been catapulted in the past from their place of origin, e.g. away
from the plane of the galaxy, and estimate the force required. If it
exceeds the force which could have been exerted by a supernova, it might
suggest an event of the sudden evaporation of a solar mass black hole
which ejected its companion. This could be taken as an indication of a
gapless CFT in the hidden sector, as in the RS2 framework. We note however
that even a slight improvement of our bound (\ref{imprbd}) would prolong
the decay time of such black holes, deferring the detonations to much
larger timescales which would make their observations much less likely.

\section{Summary and Future Directions }

In this article we have discussed phenomenological and
astrophysical aspects of black holes in the infinite
Randall-Sundrum model \cite{rs2}. Our analysis is based on the
picture proposed in \cite{efk,tanaka}, whereby a black hole larger than
the AdS radius, which is stuck on the brane, is described in the
dual CFT+gravity as a copious source of Hawking CFT radiation.
This will in general shorten the lifetime of a black hole, and
specifically, it will make large black holes lose a huge part of
their mass very quickly. We have shown that because of this effect
black holes are the most sensitive indicator of the presence of a
large number of CFT degrees of freedom in the hidden sector. A
black hole with a mass $\la M_\odot$ would impose stronger
constraints on the AdS radius $L$ than the table-top experiments
\cite{eotwash} and cosmology \cite{apr3,hmr}. Already at the
moment, X-ray binaries require that $L \la 10^{-2}$ mm. The bounds
could be as strong as $L \la {\rm few} \times 10^{-9}$ m if MACHOs
are proven to be primordial black holes. We note that
it is quite fascinating that the bounds are still fairly weak, in
spite of the fact that the ensuing number of CFT degrees of freedom
may be huge, as large as $\la 10^{60}$. This is a consequence of the
weakness of gravity and the weakness of direct couplings of CFT to the
visible sector \cite{rs2,dkkls}. We have also seen that
there remains a less likely, but exciting prospect of black hole
detonations as a signature of RS2, in the case when the AdS radius
is close to the observational bounds.

We close with some remarks on cosmological implications of black
holes in RS2 and more generally in any theory with a large number
of CFT degrees of freedom in the hidden sector in IR. Black hole
physics in such models can lead to stringent bounds on the models
of the early universe in theories with a large
number of CFT degrees of freedom. Namely the hidden sector CFT and the visible
sector (including the Standard Model) interact with each other
only very weakly, with direct couplings that are weaker than
gravitational in the infrared. Black holes provide for a stronger
channel of interaction between these two sectors. Indeed, because
CFT and CMB couple only very weakly, they can be viewed
essentially as two separate vessels of gas, insulated from each
other. A black hole acts as a pipe connecting these two vessels.
The gas can flow through the pipe from the hotter into the colder
vessel as long as there is a temperature gradient, or until the
flow of heat deteriorates the pipe, i.e. when the black hole
completely evaporates away. If the universe starts mostly
populated by the visible sector particles, ie CMB, black holes will mostly
accrete hot CMB, and will mostly evaporate colder CFT, which has
many more flavors to take away the entropy. Thus black holes will
eventually lead to the heating of the CFT sector in the universe.

On the other hand, having a realistic cosmology requires that
there was very little production of CFT radiation in the early
universe, to avoid disturbing the fine balances required for
nucleosynthesis. As shown in \cite{apr3,hmr}, the weak direct
CFT-visible sector couplings lead to a very slow production of the
CFT radiation, consistent with nucleosynthesis, as long as $L
\ltap $ 1 mm. Black holes can accelerate the conversion of the
visible sector into CFT radiation if their primordial
abundances were large. Thus requiring that the initial
abundance of primordial black holes is low enough so that the CFT
radiation remains cold enough during nucleosynthesis
will give new constraints on the early
cosmology of RS2 with a large $L$. There may
also arise bounds from structure formation, because CFT would
behave as hot dark matter (a possibility touched upon in
\cite{dkkls}). However, we refrain from a detailed investigation
of these issues here because the new bounds will depend on the
specifics of the early universe evolution and on the details of CFT
self-interactions. They may be a useful guide for building the early
universe models in RS2.

\vspace{0.5cm}
{\bf Acknowledgements}
\medskip

We gratefully acknowledge useful discussions with Jorge Casares, John
March-Russell, F\'elix Mirabel, Lisa Randall, Robert Wagoner and
Jun'ichi Yokoyama.  N.K. and J.G-B. would like to thank the Theory
Division at CERN for hospitality while this work was being initiated.
The work of R.E. has been partially supported by CICYT grants
AEN-99-0315 and FPA-2002-02037.  The work of J.G-B.  has been partially
supported by a CICYT grant FPA-2000-980. The work of N.K. has been
supported in part by UCD, in part by an award from the Research
Corporation and in part by the NSF Grant PHY-9870115 to the Stanford
ITP.

\end{document}